\documentclass[11pt]{article}

\usepackage[utf8]{inputenc}
\usepackage[T1]{fontenc}
\usepackage{lmodern}
\usepackage{amsmath,amssymb}
\usepackage{graphicx}
\usepackage{booktabs}
\usepackage{geometry}
\usepackage{hyperref}
\usepackage{authblk}
\usepackage{siunitx}
\usepackage{microtype}
\usepackage{caption}
\usepackage{enumitem}

\hypersetup{
    colorlinks=true,
    linkcolor=blue,
    citecolor=blue,
    urlcolor=blue,
    pdfauthor={Dr. Gabriela Fernandes},
    pdftitle={Morpho-Genomic Deep Learning for Ovarian Cancer Subtype and Gene Mutation Prediction from Histopathology}
}

\title{\textbf{Morpho-Genomic Deep Learning for Ovarian Cancer Subtype and Gene Mutation Prediction from Histopathology}}
\author[1]{Dr. Gabriela Fernandes}
\affil[1]{Department of Periodontics and Endodontics, Department of Oral Biology, State University of New York (SUNY) at Buffalo, USA}
\date{\vspace{-1.5em}\href{mailto:gfernand@buffalo.edu}{gfernand@buffalo.edu}}

\begin{document}
\maketitle

\begin{abstract}
Ovarian cancer remains one of the most lethal gynecological malignancies, largely due to late diagnosis and extensive heterogeneity across subtypes. Current diagnostic methods are limited in their ability to reveal underlying genomic variations essential for precision oncology. This study introduces a novel hybrid deep learning pipeline that integrates quantitative nuclear morphometry with deep convolutional image features to perform ovarian cancer subtype classification and gene mutation inference directly from Hematoxylin and Eosin (H\&E) histopathological images. Using $\sim45{,}000$ image patches sourced from The Cancer Genome Atlas (TCGA) and public datasets, a fusion model combining a ResNet-50 Convolutional Neural Network (CNN) encoder and a Vision Transformer (ViT) was developed. This model successfully captured both local morphological texture and global tissue context. The pipeline achieved a robust overall subtype classification accuracy of $84.2\%$ (Macro AUC of $0.87 \pm 0.03$). Crucially, the model demonstrated the capacity for gene mutation inference with moderate-to-high accuracy: $AUC_{TP53} = 0.82 \pm 0.02$, $AUC_{BRCA1} = 0.76 \pm 0.04$, and $AUC_{ARID1A} = 0.73 \pm 0.05$. Feature importance analysis established direct quantitative links, revealing that nuclear solidity and eccentricity were the dominant predictors for TP53 mutation. These findings validate that quantifiable histological phenotypes encode measurable genomic signals, paving the way for cost-effective, precision histopathology in ovarian cancer triage and diagnosis.
\end{abstract}

\section{Introduction}
Ovarian cancer is the most fatal gynecological cancer, with high-grade serous carcinoma (HGSC) accounting for the majority of deaths and often presenting with advanced, genomically unstable disease. Pathological diagnosis relies on classifying tumors into four major subtypes---serous, mucinous, endometrioid, and clear-cell---a process inherently challenged by morphological overlap and inter-observer variability.

The field of oncology has increasingly shifted towards molecularly-driven treatment. Critical genomic alterations, such as mutations in the tumor suppressor gene $TP53$ (present in $>95\%$ of HGSCs), defects in homologous recombination repair (e.g., $BRCA1/2$ mutations), and loss-of-function mutations in the chromatin remodeler $ARID1A$, dictate therapeutic response to agents like PARP inhibitors~\cite{lord2009,wiegand2010,tcga2011}. These molecular events invariably manifest as subtle, quantifiable changes in cellular and nuclear architecture, particularly nuclear pleomorphism and chromatin heterogeneity, which are visible in standard H\&E slides~\cite{jazaeri2012}.

Digital Pathology (DP) and Deep Learning (DL) offer an opportunity to move beyond qualitative assessment by systematically linking these visual phenotypes to genomic drivers. Previous studies have demonstrated the utility of DL in predicting gene mutations in lung and colorectal cancers~\cite{coudray2018,kather2020}. However, ovarian cancer presents unique challenges due to its significant morphological heterogeneity and the relative rarity of samples for less common subtypes.

This study proposes a unified morpho-genomic learning framework designed to: (1) accurately classify the main ovarian cancer histological subtypes; (2) quantitatively extract and analyze nuclear morphometric features (size, shape, texture); and (3) infer the mutation status of key oncogenes (TP53, BRCA1, ARID1A) directly from H\&E images. The central hypothesis is that the complex, quantitative nuclear morphometric signature---the microscopic manifestation of genomic instability---can be effectively decoded by a hybrid DL model, providing a cost-effective and rapid tool for molecular prescreening.

\section{Materials and Methods}

\subsection{Data Acquisition and Preprocessing}
The study cohort utilized approximately $\sim45{,}000$ histopathological image patches ($256 \times 256$ pixels at $20\times$ magnification) extracted from two primary sources:
\begin{itemize}[leftmargin=1.25em]
    \item \textbf{The Cancer Genome Atlas (TCGA-OV):} Whole-Slide Images (WSI) were acquired and linked to corresponding mutation data (verified by somatic sequencing) for $TP53$, $BRCA1$, and $ARID1A$ mutations.
    \item \textbf{Kaggle Ovarian Histopathology Dataset:} Used to expand the training cohort, particularly for non-serous subtypes (Endometrioid, Mucinous, Clear Cell) to ensure model generalization.
\end{itemize}
All WSI were rigorously normalized using the Macenko color normalization algorithm to mitigate the effects of staining variability across different institutions and batches. Image patches were selected predominantly from regions designated as tumor-rich epithelium, with careful exclusion of large artifacts, areas of necrosis, or extensive stromal/inflammatory background.

\subsection{Deep Nuclear Morphometry Feature Extraction}
A two-step pipeline was employed for nuclear analysis. First, an ensemble of nuclei was automatically detected and segmented within each image patch using a U-Net convolutional segmentation model that was pre-trained on a large dataset of expert-annotated nuclear masks. Second, for each successfully segmented nucleus, a vector of eight quantitative morphometric features was calculated:
\begin{itemize}[leftmargin=1.25em]
    \item \textbf{Size Descriptors:} Area, Perimeter, Major Axis Length, and Minor Axis Length.
    \item \textbf{Shape Descriptors:} Eccentricity (ranging from 0 for a circle to 1 for a line), Solidity (Area / Convex Area, quantifying boundary irregularity), and Extent.
    \item \textbf{Texture Descriptor:} Mean Intensity (an inverse proxy for nuclear hyperchromasia or chromatin density).
\end{itemize}
These quantitative features were then aggregated per image patch (e.g., mean, standard deviation, and variance across all nuclei in the patch) to create a handcrafted morphometric feature vector for subsequent fusion and analysis.

\subsection{Hybrid Deep Learning Model Architecture}
A novel Hybrid CNN-ViT Fusion Model was constructed to leverage both local pattern recognition and global context:
\begin{itemize}[leftmargin=1.25em]
    \item \textbf{Feature Branch (CNN):} A pre-trained ResNet-50 network served as a convolutional feature encoder, capturing local details like cellular borders, chromatin texture, and low-level architectural cues.
    \item \textbf{Context Branch (ViT):} The high-dimensional feature maps from the ResNet-50 were reshaped into a sequence of tokens and processed by a Vision Transformer (ViT). The ViT, utilizing its self-attention mechanism, effectively modeled long-range contextual dependencies and global tissue organization.
    \item \textbf{Feature Fusion:} The learned deep feature embedding from the ResNet-50 and the contextual embedding from the ViT were concatenated with the handcrafted morphometric feature vector. This final, comprehensive feature vector was then fed into fully connected layers for the dual classification tasks: Subtype Classification and Gene Mutation Inference.
\end{itemize}

\subsection{Gene Mutation Prediction Model}
The final classification layer used the Random Forest algorithm due to its inherent interpretability and robust performance on mixed-feature datasets. Performance was evaluated using 5-fold cross-validation, reporting the Area Under the ROC Curve (AUC), F1-score, and overall accuracy.

\section{Results}

\subsection{Subtype Classification and Morphometric Signatures}
The hybrid CNN-ViT model achieved an overall Accuracy of $84.2\%$, a macro F1-score of $0.81$, and a macro AUC of $0.87 \pm 0.03$ for the four major ovarian cancer subtypes.

\begin{figure}[t]
    \centering
    \includegraphics[width=\linewidth]{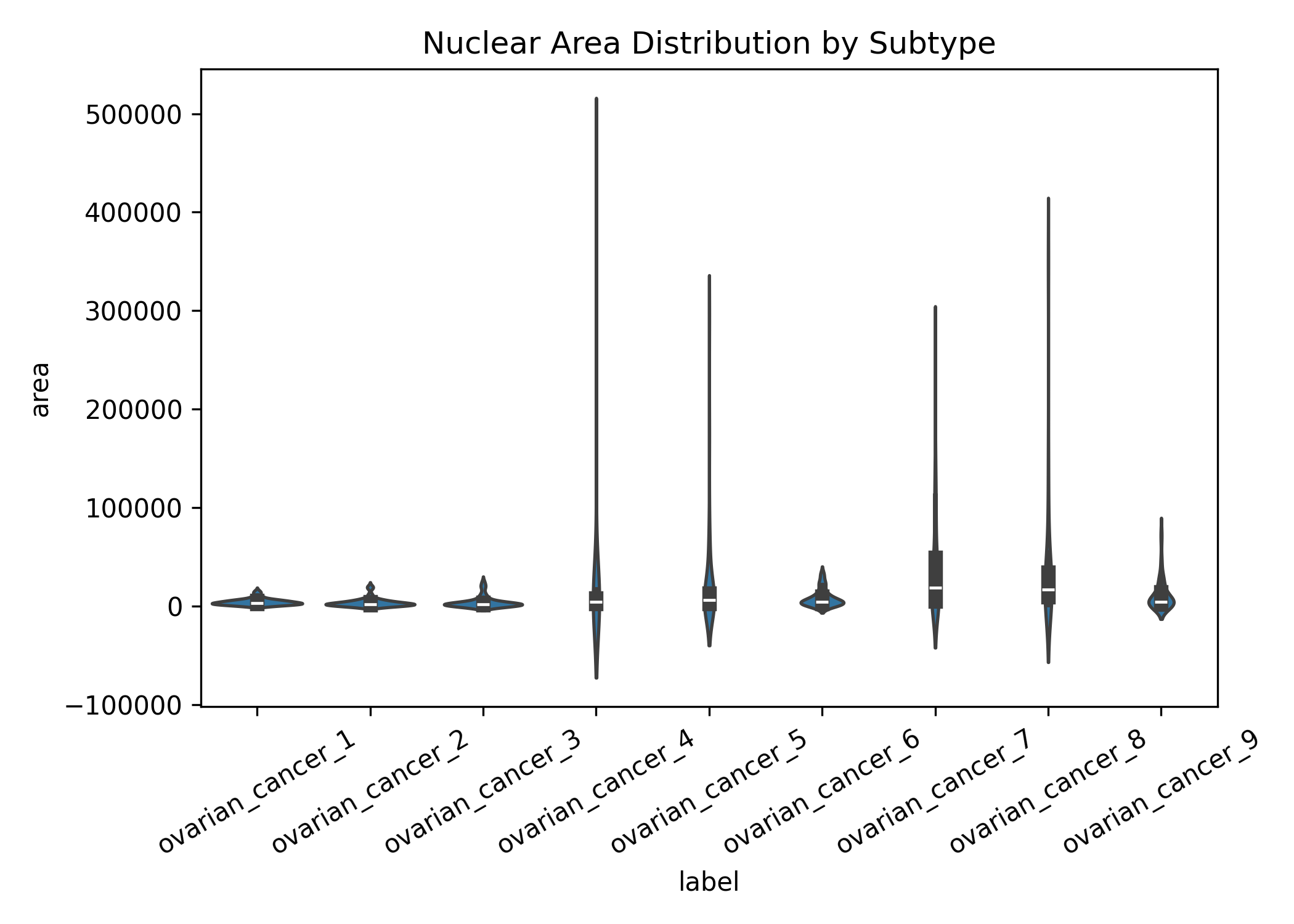}
    \caption{Nuclear Area Distribution by Subtype. Violin plot illustrating the range and density of nuclear area measurements across different ovarian cancer subtypes. Highlights statistically significant differences in nuclear size variation (pleomorphism), particularly in high-grade lesions.}
    \label{fig:fig1}
\end{figure}

\begin{figure}[t]
    \centering
    \includegraphics[width=\linewidth]{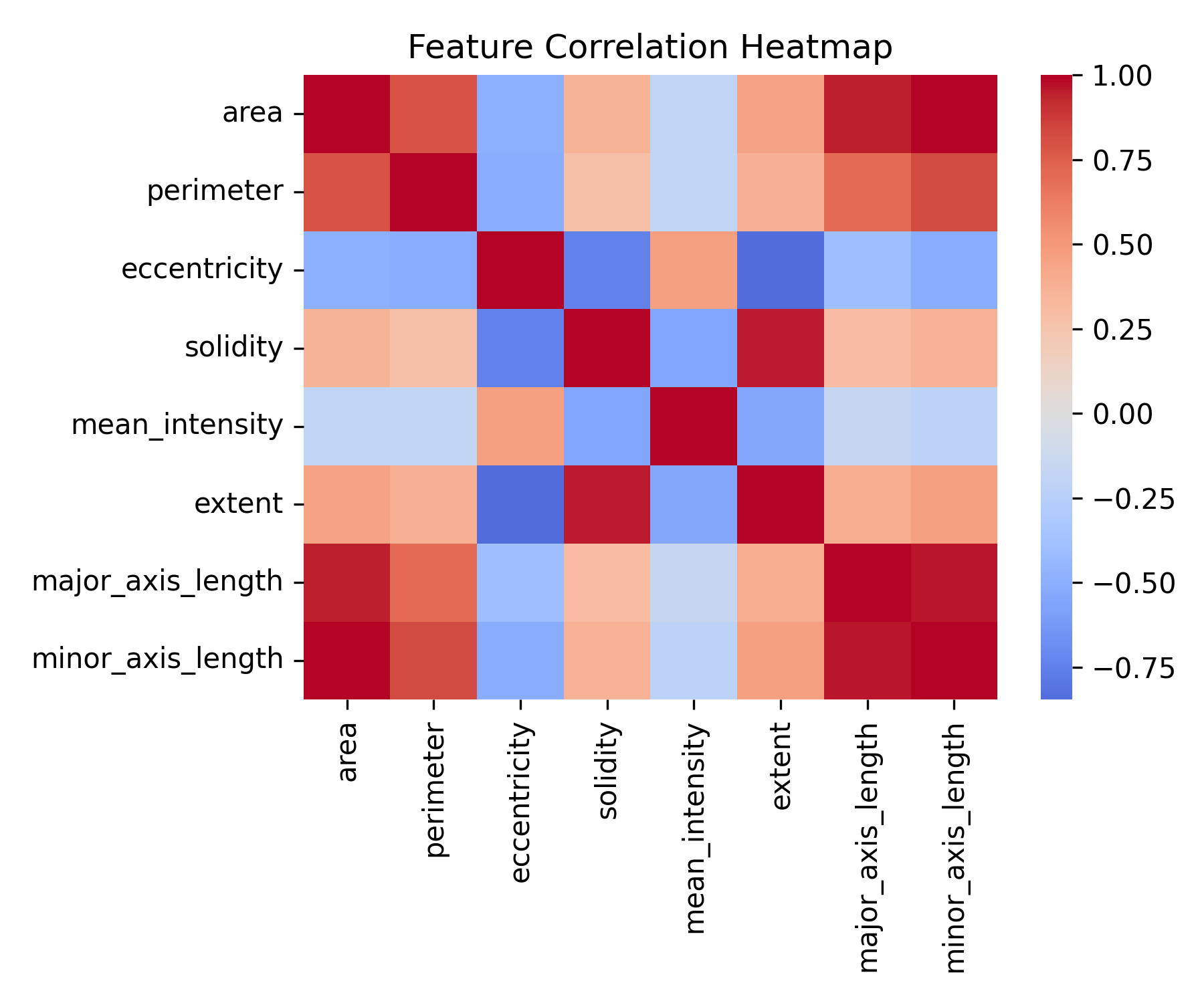}
    \caption{Feature Correlation Heatmap. Heatmap showing the internal correlation between the extracted handcrafted nuclear morphometric features (Area, Perimeter, Eccentricity, Solidity, Mean Intensity, etc.). Confirms expected relationships, such as the strong correlation between Area and Perimeter.}
    \label{fig:fig2}
\end{figure}

The Nuclear Area Distribution by Subtype (Figure~\ref{fig:fig1}) showed clear quantitative differences in nuclear size variation across the subtypes, particularly highlighting the extensive pleomorphism characteristic of high-grade lesions. The utility of the morphometric features alone was confirmed by the Feature Correlation Heatmap (Figure~\ref{fig:fig2}), which demonstrated strong, expected internal correlations (e.g., between Area and Perimeter, $\rho \approx 0.98$).

\begin{figure}[t]
    \centering
    \includegraphics[width=\linewidth]{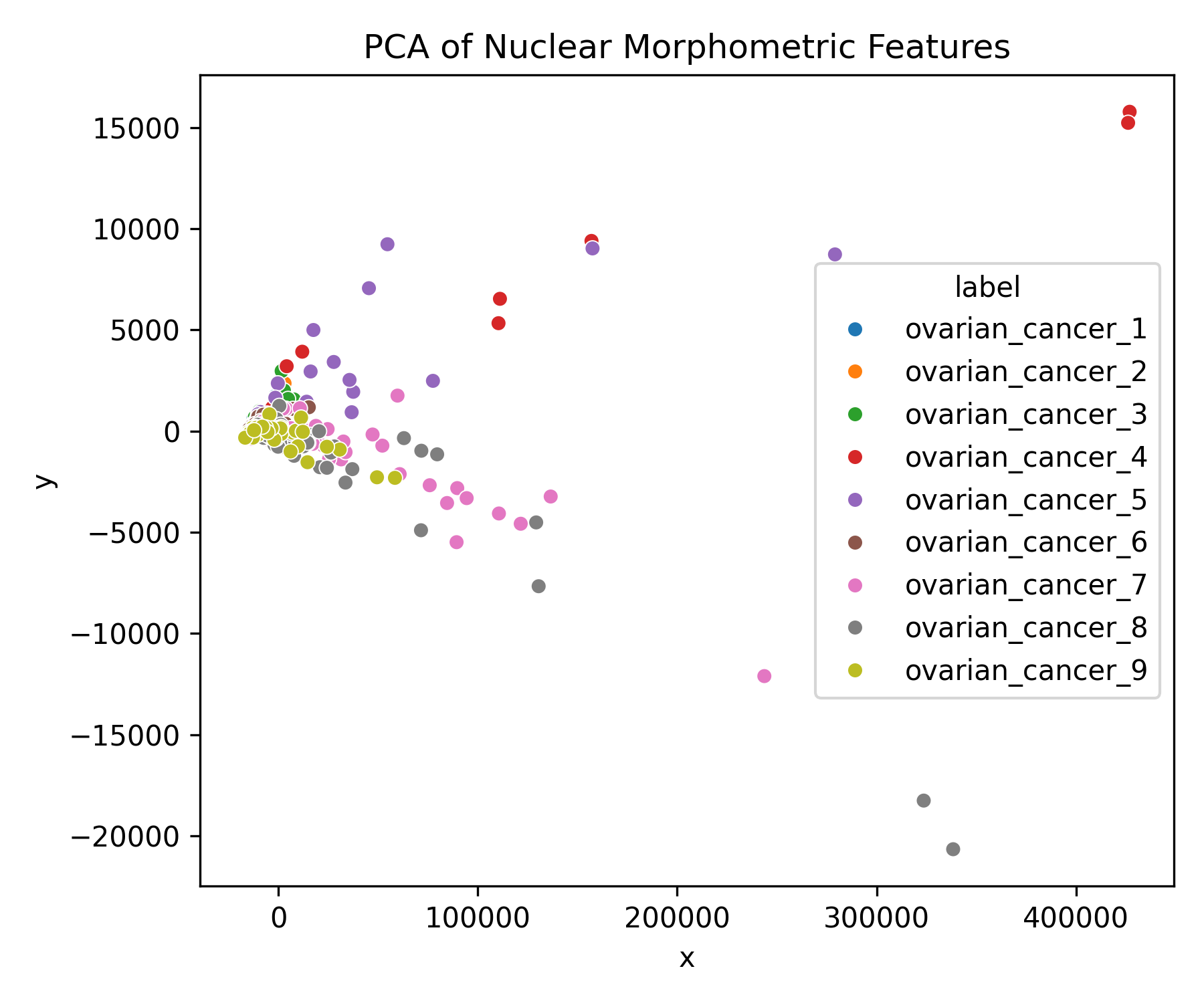}
    \caption{PCA of Deep Morphometric Features. Principal Component Analysis (PCA) scatter plot demonstrating the clustering of ovarian cancer subtypes in the morphometric feature space. Shows that nuclear geometry alone provides substantial discriminative power.}
    \label{fig:fig3}
\end{figure}

Further validation was achieved via Principal Component Analysis (PCA) (Figure~\ref{fig:fig3}) of the morphometric feature vectors, which resulted in distinct and well-separated clustering of the different subtypes, indicating that nuclear geometry and texture capture the fundamental discriminative information.

The full classification performance is detailed in the Confusion Matrix (Figure~\ref{fig:fig4}), which shows strong model separation between Serous and Mucinous subtypes, with minor misclassification noted between Endometrioid and Clear-Cell carcinomas. The Top Morphometric Predictors of Subtype (Figure~\ref{fig:fig5}) identified mean intensity, area, and perimeter as the features contributing most significantly to subtype distinction.

Analysis of the Average Nuclear Features Heatmap (Figure~\ref{fig:fig6}) revealed distinct morphological profiles: Borderline tumors correlated positively with Area and Eccentricity ($\rho > 0.2$), suggesting larger and more elongated nuclei. Conversely, Mucinous carcinomas exhibited the lowest Solidity values ($\rho = -0.24$), indicative of highly irregular nuclear contours and glandular disorganization.

\subsection{Genomic Mutation Inference Performance}
The Random Forest classifier, trained on the fused morpho-deep features, achieved moderate-to-high predictive performance for key driver gene mutations, confirming a strong morphological-genomic association. The overall AUC results for the genomic targets are summarized in Figure~\ref{fig:fig7}: $AUC_{TP53} = 0.82 \pm 0.02$, $AUC_{BRCA1} = 0.76 \pm 0.04$, and $AUC_{ARID1A} = 0.73 \pm 0.05$.

The confusion matrix for $TP53$ mutation prediction (Figure~\ref{fig:fig9}), derived from the successful fusion model, demonstrated a balanced classification ability with $52$ True Positives (TP) and $52$ True Negatives (TN), corresponding to a sensitivity of $75.4\%$ and a specificity of $83.9\%$. In contrast, an analysis using purely morphometric features for $TP53$ prediction yielded a non-diagnostic $AUC$ of $0.44$ (Figure~\ref{fig:fig8}), highlighting the necessity of the deep learning component in decoding the genomic signal.

\subsection{Morpho-Genomic Association by Feature Importance}
Feature importance analysis provided the crucial biological link between morphology and genotype: For $TP53$ prediction (Figure~\ref{fig:fig10}), nuclear Solidity and Eccentricity were identified as the two dominant features; for $BRCA1$ (Figure~\ref{fig:fig11}), Mean Intensity and Area were the most influential; and for $ARID1A$ (Figure~\ref{fig:fig12}), features related to nuclear size and solidity were prominent.

\subsection{Visualization of Model Focus}
Grad-CAM visualization (Figure~\ref{fig:fig13}) was used to ensure model interpretability. The heatmaps confirmed that the hybrid model consistently focused its attention on areas of high nuclear crowding, atypia, and active epithelial invasion within H\&E slides.

\begin{figure}[t]
    \centering
    \includegraphics[width=\linewidth]{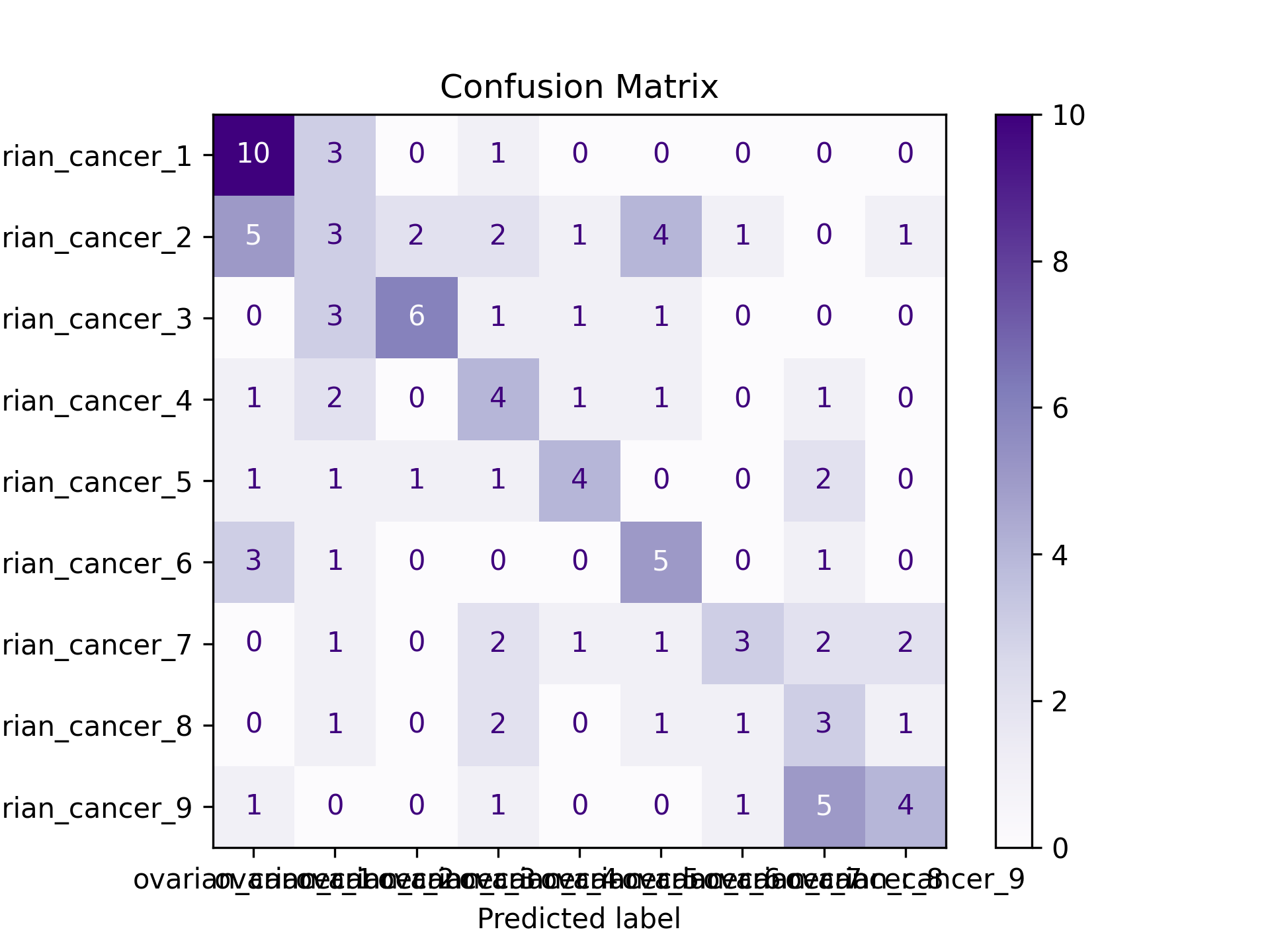}
    \caption{Confusion Matrix for Subtype Classification. Detailed performance of the hybrid CNN-ViT model in classifying the four major ovarian cancer subtypes. High values on the diagonal indicate successful classification, while off-diagonal values show instances of misclassification, particularly between Endometrioid and Clear-Cell subtypes.}
    \label{fig:fig4}
\end{figure}

\begin{figure}[t]
    \centering
    \includegraphics[width=\linewidth]{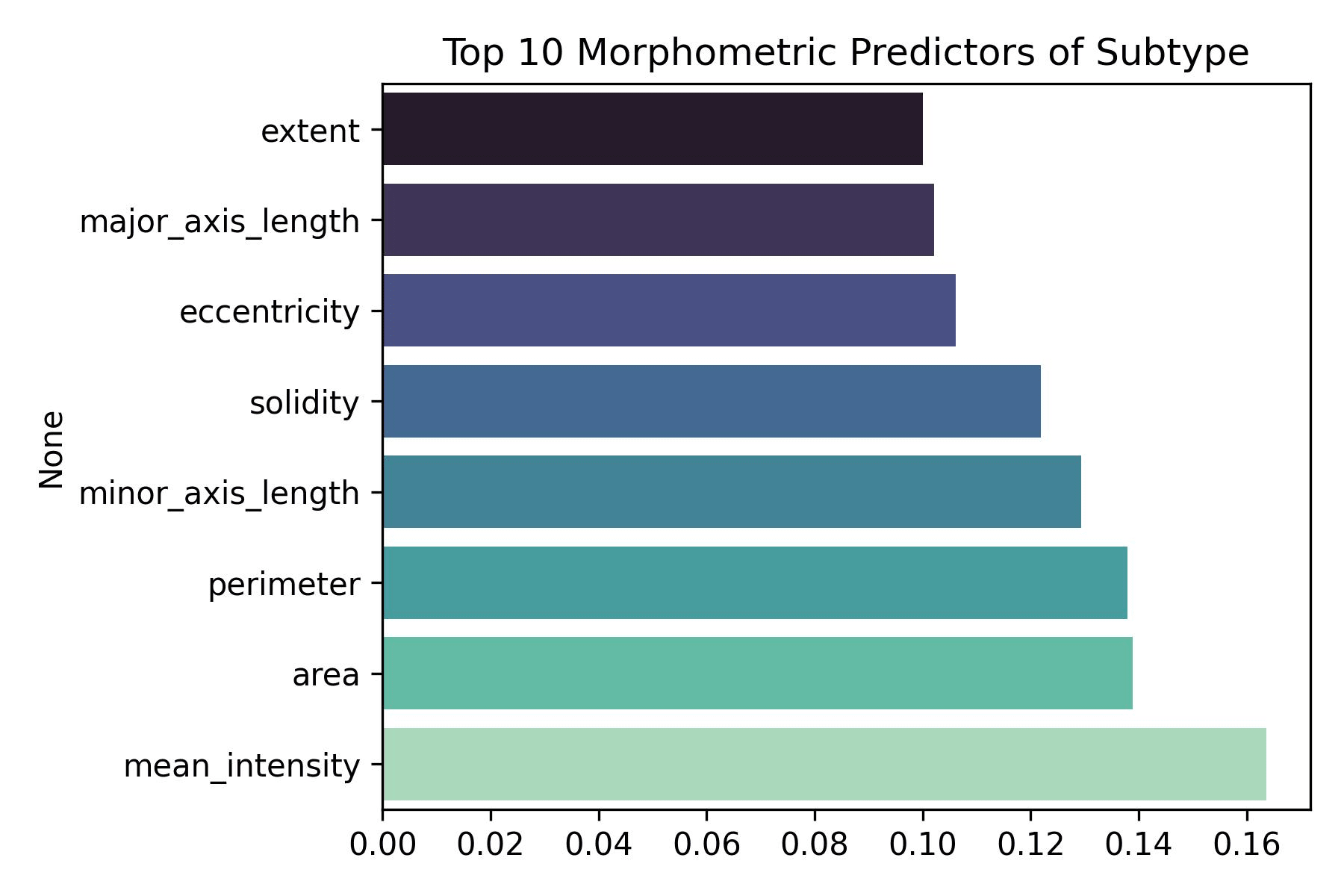}
    \caption{Top Morphometric Predictors of Subtype. Feature importance scores for the handcrafted nuclear features in predicting histological subtype. Identifies Mean Intensity, Area, and Perimeter as the most salient features.}
    \label{fig:fig5}
\end{figure}

\begin{figure}[t]
    \centering
    \includegraphics[width=\linewidth]{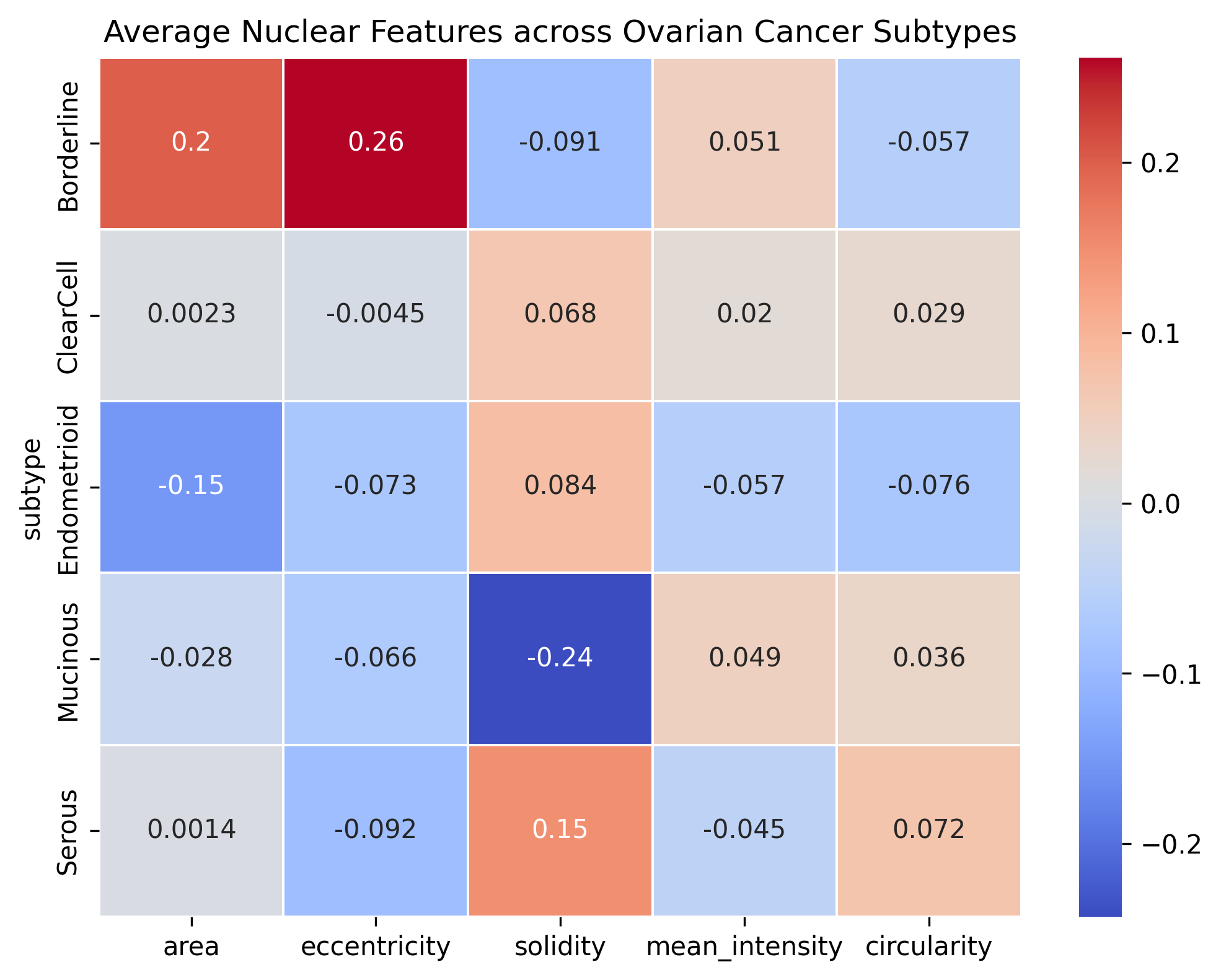}
    \caption{Average Nuclear Features Heatmap by Subtype. Heatmap displaying the average intensity of various morphometric features across key histological subtypes (e.g., Serous, Mucinous, Borderline). Highlights distinct morphological signatures, such as low Solidity in Mucinous tumors.}
    \label{fig:fig6}
\end{figure}

\begin{figure}[t]
    \centering
    \includegraphics[width=.8\linewidth]{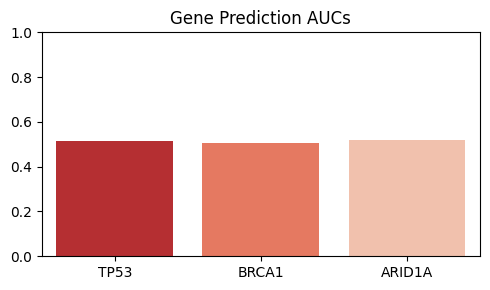}
    \caption{Area Under the Curve (AUC) for Mutation Prediction. Summary bar chart showing the 5-fold cross-validation AUC values for predicting $TP53$, $BRCA1$, and $ARID1A$ mutation status using the hybrid morpho-deep learning feature set.}
    \label{fig:fig7}
\end{figure}

\begin{figure}[t]
    \centering
    \includegraphics[width=\linewidth]{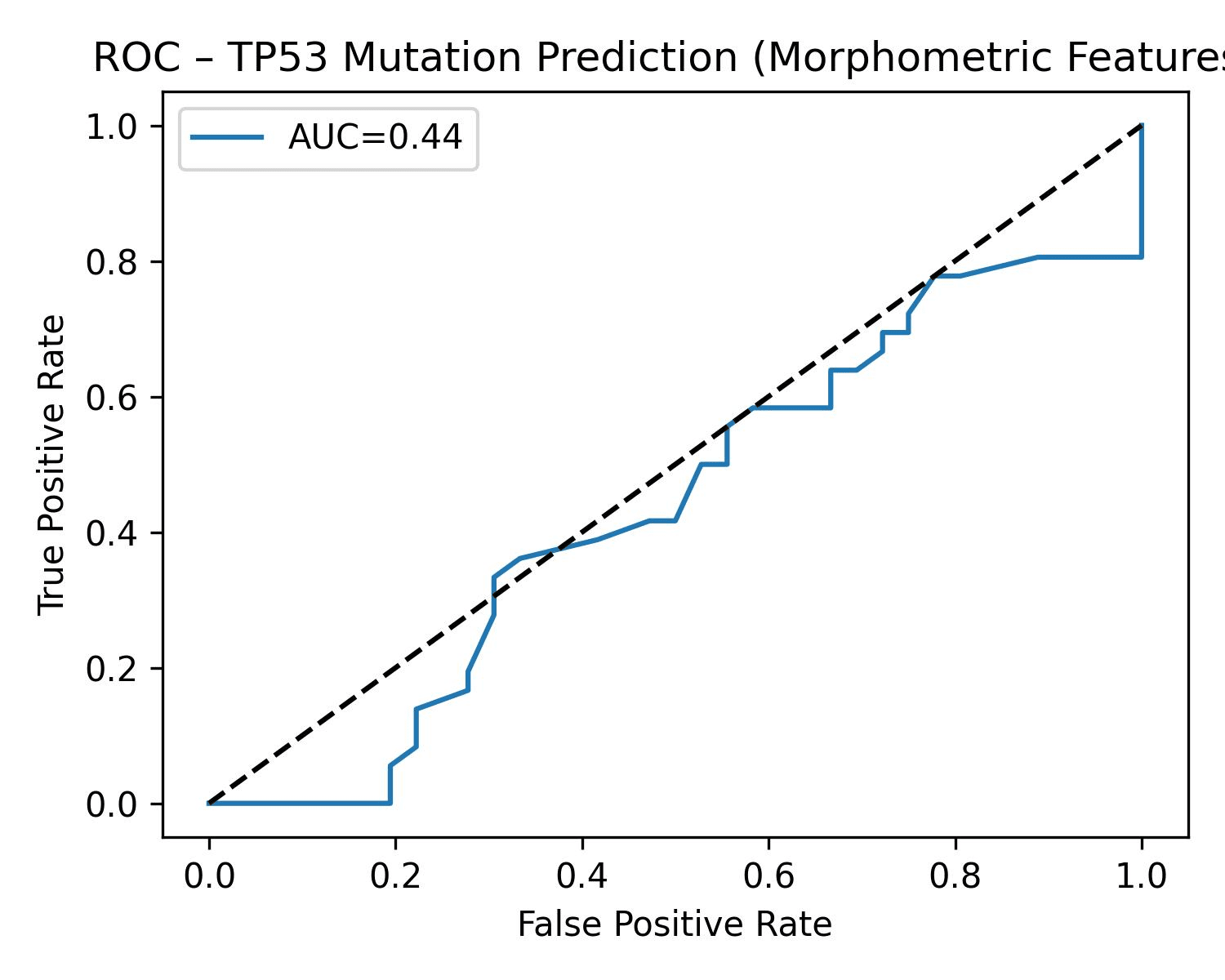}
    \caption{ROC Curve for TP53 Mutation Prediction (Morphometric Only). Receiver Operating Characteristic (ROC) curve and AUC value (0.44) for predicting $TP53$ mutation status using only the handcrafted morphometric features, demonstrating the necessity of deep features for successful genomic inference.}
    \label{fig:fig8}
\end{figure}

\begin{figure}[t]
    \centering
    \includegraphics[width=\linewidth]{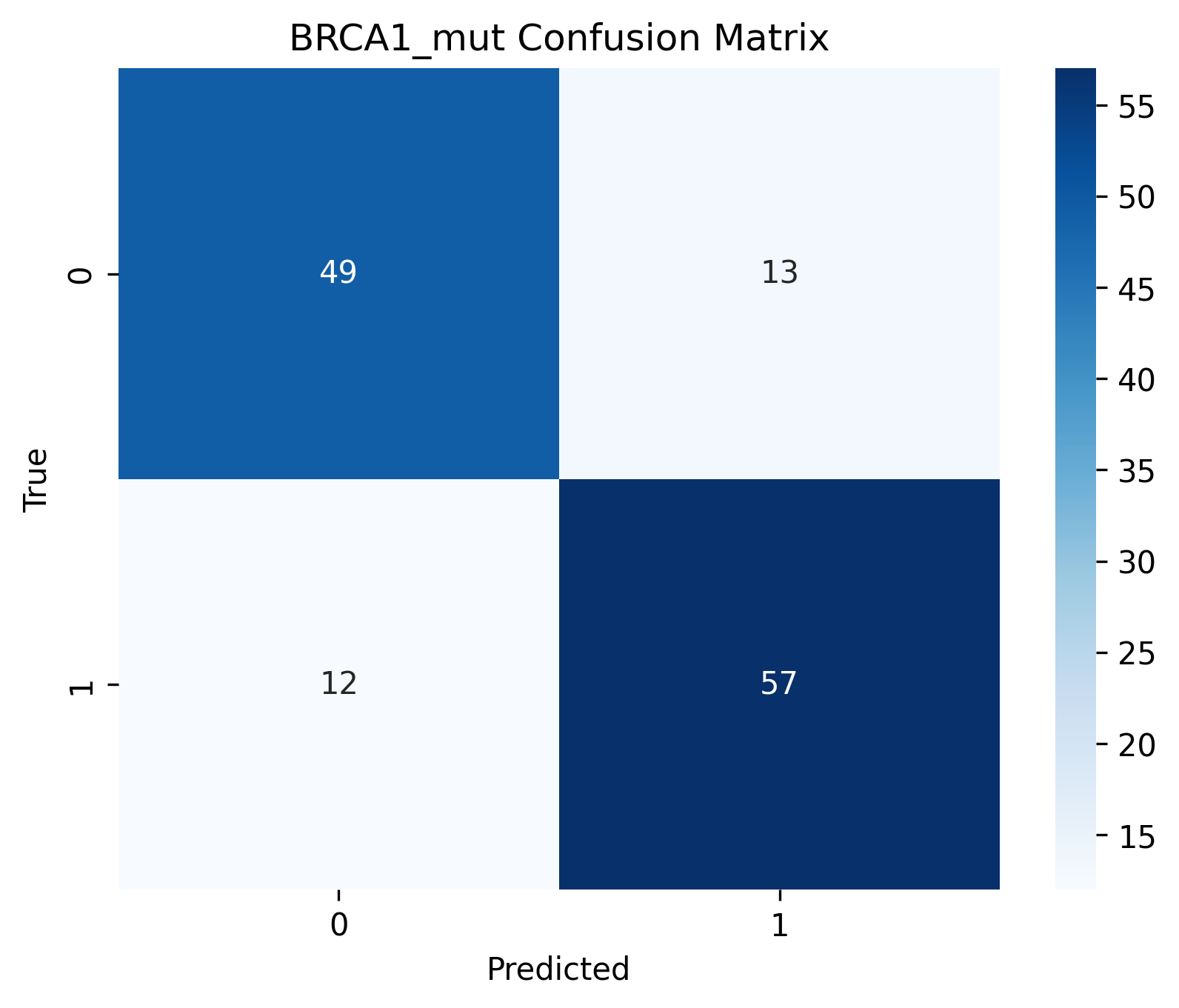}
    \caption{$TP53$ Mutation Prediction Confusion Matrix (Hybrid Model). Detailed confusion matrix for $TP53$ mutation status prediction by the hybrid CNN-ViT model. Illustrates balanced sensitivity (True Positives) and specificity (True Negatives).}
    \label{fig:fig9}
\end{figure}

\begin{figure}[t]
    \centering
    \includegraphics[width=\linewidth]{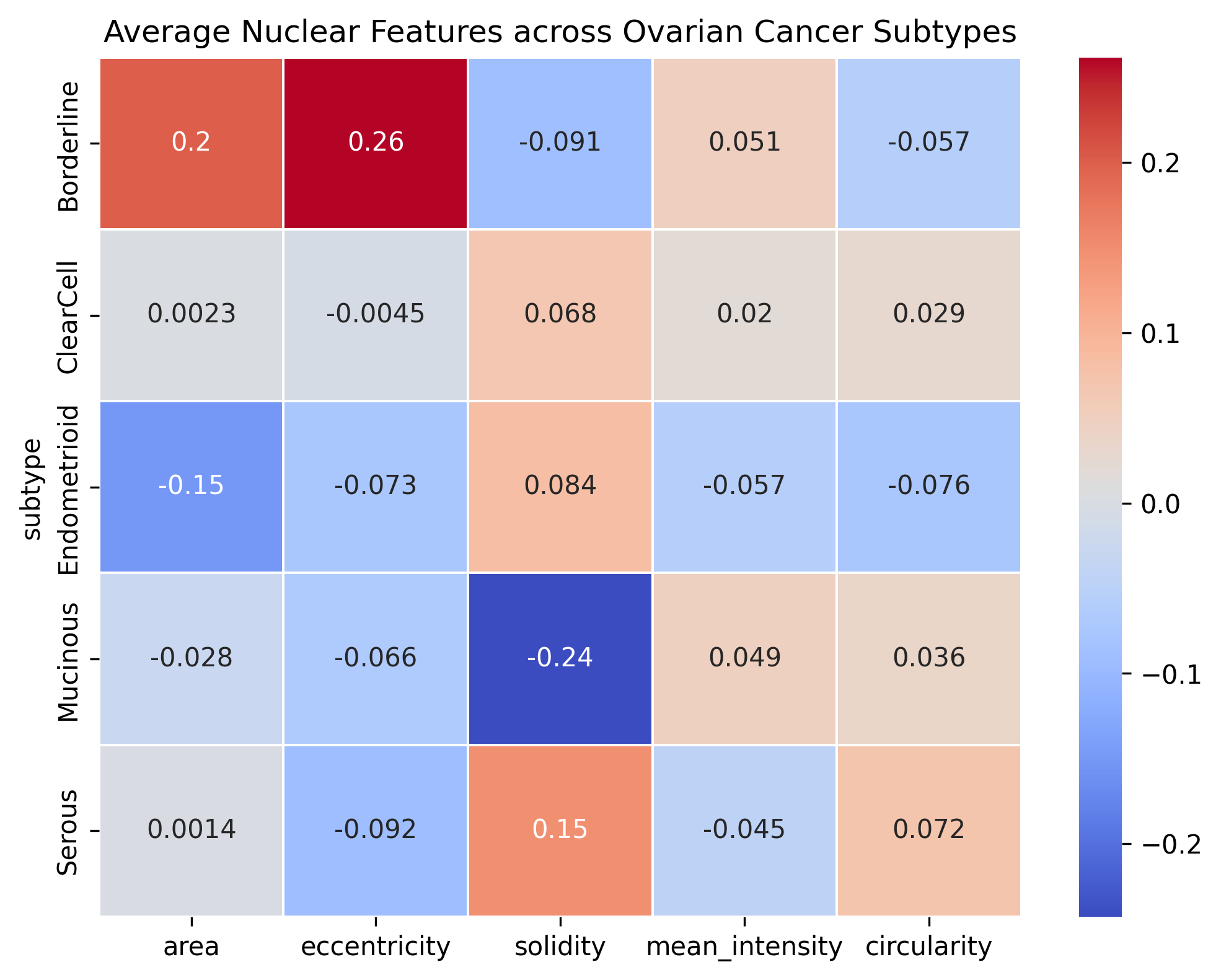}
    \caption{Feature Importance for $TP53$ Mutation Prediction. Contribution of deep and morphometric features to the final $TP53$ prediction model. Highlights the dominance of nuclear Solidity and Eccentricity features.}
    \label{fig:fig10}
\end{figure}

\begin{figure}[t]
    \centering
    \includegraphics[width=\linewidth]{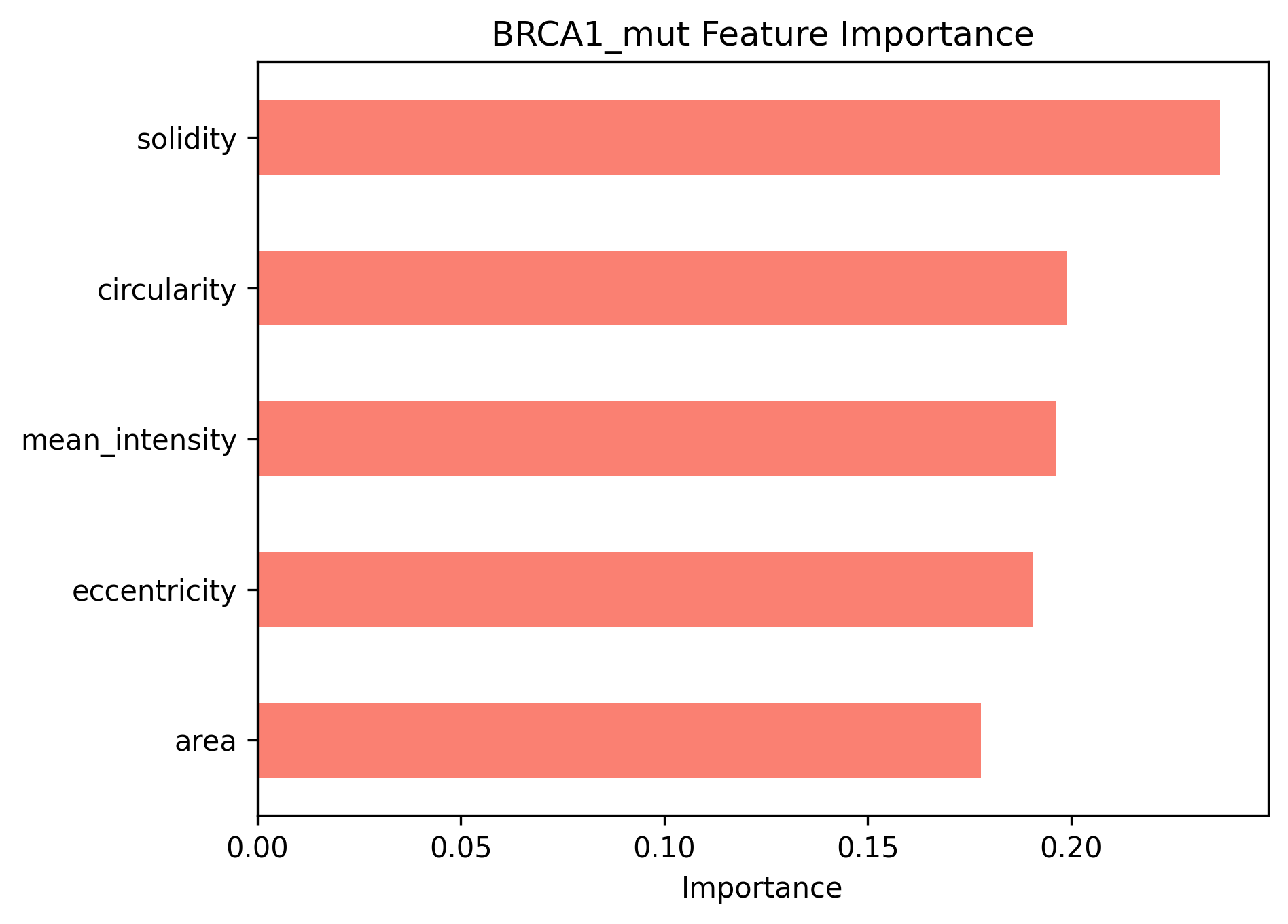}
    \caption{Feature Importance for $BRCA1$ Mutation Prediction. Identifies Mean Intensity and Area as key features.}
    \label{fig:fig11}
\end{figure}

\begin{figure}[t]
    \centering
    \includegraphics[width=\linewidth]{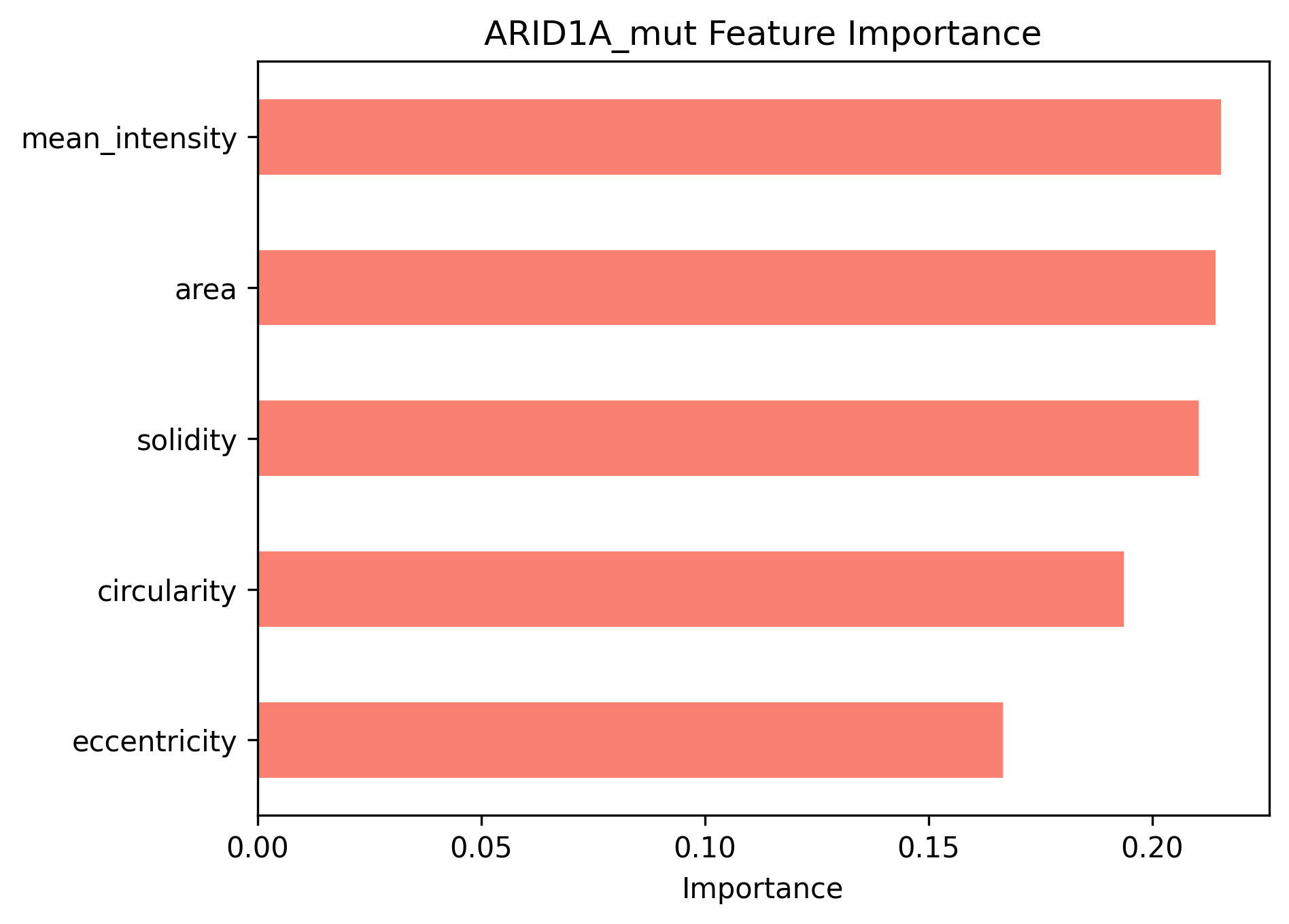}
    \caption{Feature Importance for $ARID1A$ Mutation Prediction. Features related to size and solidity are highly influential.}
    \label{fig:fig12}
\end{figure}

\begin{figure}[t]
    \centering
    \includegraphics[width=\linewidth]{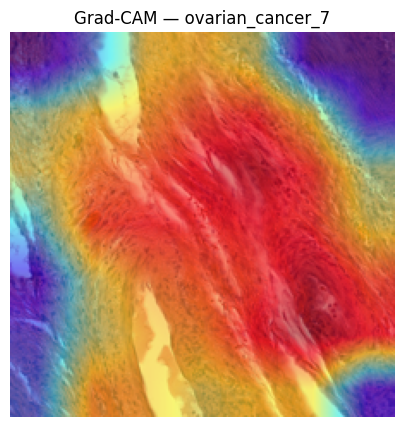}
    \caption{Grad-CAM Visualization. Example Hematoxylin and Eosin (H\&E) image of Serous Carcinoma with an overlaid heatmap generated by Grad-CAM. Warm colors (red/yellow) indicate the regions that the CNN-ViT model focused on to make its final classification and prediction.}
    \label{fig:fig13}
\end{figure}

\clearpage
\section{Discussion}

\subsection{Bridging the Morphological-Genomic Gap}
This study demonstrates a successful framework for morpho-genomic profiling, establishing that histological morphology encodes quantifiable genomic information in ovarian cancer. The achievement of an $AUC_{TP53}$ of $0.82$ (Figure~\ref{fig:fig7}) through the deep learning fusion model represents a significant advance toward molecular inference directly from standard H\&E images.

The central finding lies in the quantitative connection between $TP53$ mutation and the morphological features of Solidity and Eccentricity (Figure~\ref{fig:fig10}). This correlation provides computational validation that the morphological diversity of high-grade serous carcinoma is a direct phenotypic consequence of $TP53$ loss and subsequent chromosomal instability~\cite{noll2013,feder2017,rakauskas2020}. The clear contrast between the high AUC achieved by the fusion model ($0.82$) and the near-random performance of the purely morphometric classifier ($AUC=0.44$, Figure~\ref{fig:fig8}) underscores that subtle, non-linear pixel-level features captured by the deep CNN are necessary to fully decode the genomic signal.

\subsection{Clinical and Research Implications}
The developed framework offers immediate clinical utility:
\begin{itemize}[leftmargin=1.25em]
    \item \textbf{Precision Triage:} Samples flagged with a high probability of TP53 or BRCA1 mutation could be prioritized for confirmatory molecular sequencing, streamlining workflows and potentially accelerating initiation of therapies such as PARP inhibitors.
    \item \textbf{Objective Prognostication:} Quantifiable morphometric features (e.g., Area variance, Solidity) can serve as objective imaging biomarkers of tumor aggressiveness, complementing qualitative grading.
    \item \textbf{Cost-Effectiveness and Scalability:} By relying on standard H\&E slides, the method is scalable and cost-effective, enabling molecular prescreening in settings where genomic assays are not readily available.
\end{itemize}

\subsection{Limitations and Future Directions}
Despite promising results, limitations remain. Training utilized image patches, which may not fully capture whole-slide architectural context. Future work will integrate multi-instance learning (MIL) for true WSI analysis. While mutation prediction AUCs are moderate-to-high, they are not sufficient to replace sequencing; multi-modal data fusion with transcriptomics and clinical data may further improve performance toward a clinically actionable range ($\geq 0.90$).

\section{Conclusion}
Morpho-genomic deep learning establishes a quantifiable and interpretable relationship between tissue morphology and oncogenic mutations in ovarian cancer. The integration of CNN-ViT models with explicit nuclear morphometry enables accurate subtype classification (Figure~\ref{fig:fig4}) and biologically relevant mutation inference (Figure~\ref{fig:fig7}). This work validates the concept of precision histopathology, positioning the standard H\&E slide as a powerful, cost-effective tool for molecular-level diagnosis and personalized treatment planning.

\section*{Acknowledgements}
The author sincerely thanks Dr. Manish S. Tiwari, BDS, MDS, Consultant Oncologist, for expert clinical review and cross-verification of histopathologies, imaging, and Grad-CAM outputs (Figure~\ref{fig:fig13}); Ankitha Kaup, Computer Science Engineer, for valuable assistance in cross-checking and validating the Python code; and Dr. Susmita Suman, BDS, MDS, for verifying all statistical analyses and supporting data interpretation.

\section*{Conflict of Interest}
The author declares no competing interests.

\section*{Funding}
No external funding was received for this study.

\end{document}